# A classical to quantum optical network link for orbital angular momentum carrying light

Zhi-Yuan Zhou,[1,2] Yan Li,[1,2] Dong-Sheng Ding,[1,2] Wei Zhang,[1,2] Shuai Shi,[1,2] Bao-Sen Shi,[1,2*] and Guang-Can Guo[1,2]

[1]*Key Laboratory of Quantum Information, University of Science and Technology of China, Hefei, Anhui 230026, China*

[2]*Synergetic Innovation Center of Quantum Information & Quantum Physics, University of Science and Technology of China, Hefei, Anhui 230026, China*

[*]*Corresponding author: drshi@ustc.edu.cn*

Light with orbital angular momentum (OAM) has great potentials in both classical and quantum optical communications such as enhancing the transmission capacity of a single communication channel because of its unlimited dimensions. Based on OAM conservation in second order nonlinear interaction processes, we create a classical to quantum optical network link in OAM degree of freedoms of light via sum frequency generation (SFG) following by a spontaneous parametric down conversion (SPDC). A coherent OAM-carrying beams at telecom wavelength 1550nm is up-converted to 525.5nm OAM-carrying beams in the first crystal, then up-converted OAM-carrying beam is used to pump a second crystal to generate non-degenerate OAM entangled photon pairs at 795nm and 1550nm. By switching the OAM carries by the classical party, the OAM correlation in the quantum party is shifted. High OAM entanglements in two dimensional subspaces are verified. This primary study enables to build a hybrid optical communication network contains both classical and quantum optical network nodes.

The field of optical communication consists of two rather different schemes, classical optical communication and quantum optical communications, these two schemes differ in the principles of physical realizations such as methods of encoding, decoding and detection [1]. The information transmission capacity and security are two most important indexes to estimate the quality an optical communication network. To enhance the transmission capacity of a single communication channel, multiplexing of different degree of freedoms of light are used. In the early stages, time multiplexing, wavelength multiplexing and polarization multiplexing have been used to dramatically increasing the communication speed. Recently, multiplexing the spatial degree of freedoms of light has been demonstrated both in free space [2, 3] and fiber [4] optical communications for orbital angular momentum (OAM) carried light beam, the achieved transmission rate has reached Tbit/s. To pursue unconditional security in optical communication, quantum optical communication is developed since the born of quantum mechanics. Based on the basic principles of quantum mechanics, quantum communication is unconditional security in principle. Encoding information in OAM degree of freedoms of photon can enhance both the information carrying capacity and tolerance of noise in quantum key distribution [5-7].

Since first introduced by L. Allen et. al. [8], OAM-carried light has been widely studied in many fields such as optical trapping and manipulation [9, 10], metrology [11-13], test the basic principle of quantum mechanics [14-17] and optical communications [2-4, 7]. The total OAM of a light beam is conserved in nonlinear optical interaction processes (second order [18-22], third order [23-25] or higher order [26]). The conservation of OAM in nonlinear interaction processes is very useful for frequency conversion of OAM-carrying beams[18-22] and creating OAM entangled photon pairs in 2-dimensional subspaces[27-31] or higher dimensions [16, 32]. Because of this conservation law, it is possible to create a link between classical optical network and quantum optical network encoded in OAM degree of freedoms of light. This link acts as a compatible interface to join the classical one and quantum one into a hybrid network, which will enable information to transfer between two complete different communication systems.

In this article, we report the building of a hybrid optical network by link the classical network and quantum network using OAM degree of freedoms of light beam based on sum frequency generation (SFG) followed by spontaneous parametric down conversion (SPDC) in nonlinear crystals. The coherent OAM-carrying beam at telecom wavelength 1550nm is efficiently up-converted to 525.5nm OAM-carrying beam in the first periodically poled KTP (PPKTP1) crystals, then the up-converted OAM-carrying beam is used to pump crystal PPKTP2 to generate non-degenerate OAM entangled photon pairs at 795nm and 1550nm. Since the total OAM is conserved in this cascade processes, by switching the OAM carried by the classical party, the OAM correlation of quantum party is shifted. The dependence of the quantum OAM correlation on the OAM carried by the classical beam implies information is transfer from the classical party to the quantum party in OAM degree of freedoms of light beams. In addition, high OAM entanglements in 2-dimensional subspaces are verified via different methods: interference, CHSH inequality and quantum state tomography.

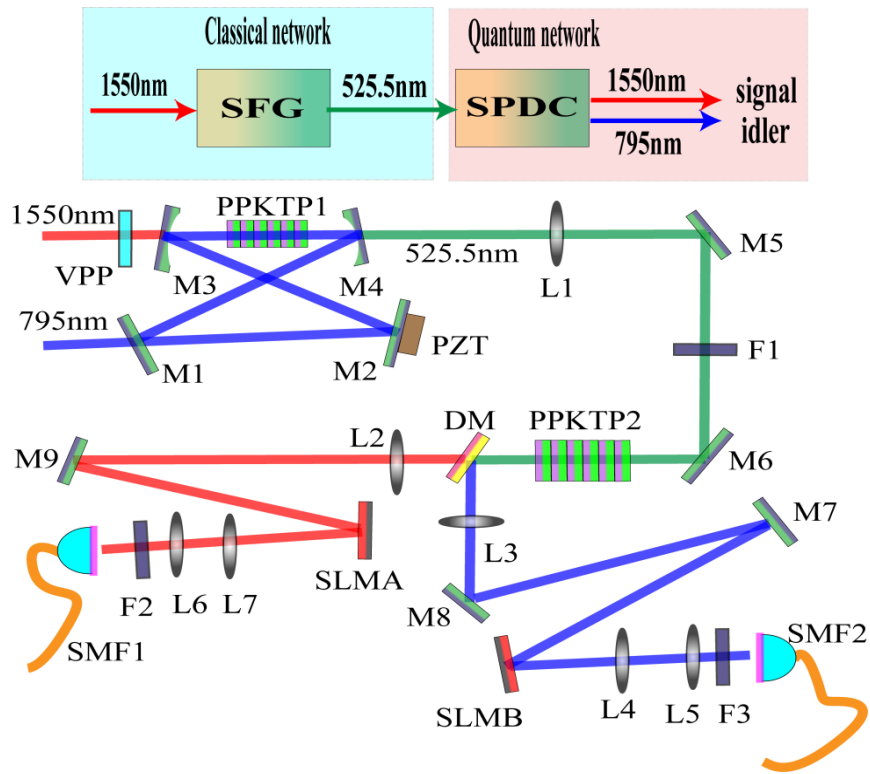

Figure1. Experimental setup for our experiment. The top simplified diagram shows the information transferring direction. M1-M9: mirrors; VPP: vortex phase plate; F1-F3: filters; DM: dichromatic mirror; PPKTP1, 2: periodically poled KTP crystals; SMF1, 2: single mode fibers; L1-L3:150mm lenses; L4, L7:200mm lenses; L5, L6:100mm lenses; PZT: piezoelectric transducer; SLMA, B: spatial light modulators.

**Results**

The experiment consists of two parts, the first part is efficiently up-conversion of OAM carried light beam based on SFG in an external cavity, the second part is SPDC using the up-converted OAM beams to generate non-degenerate OAM entangled photon pairs (see the simple diagram in the top part of Figure 1). High efficiency frequency conversion of OAM-carried light beams in periodically poled nonlinear crystals has been study in detail by us in refs. [19-22]. In the present experiment (Figure 1), the external cavity is the same as in ref. [22]. The 1550nm beam's power is 12mW (Toptica prodesign), the 795nm beam's power is 700mW (Ti:sapphire, MBR110). The power of the up-converted 525.5nm beam for OAM indexes of 0 and -1 are 0.7mW and 0.15mW after removing the two pump beams using filters F1 (Thorlabs, FBH520-40, FESH1000), respectively. The up-converted OAM-carrying beam is imaged into another crystal PPKTP2 using 150mm lens (L1) for SPDC with a magnification of about 2, therefore the beam diameter at the center of PPKTP2 is about 130 μm. Crystal PPKTP2 emits non-degenerate photon pairs at 795nm and 1550nm. The two photons are separated by a dichromatic mirror (DM), then they are imaged to spatial light modulator A and B (SLMA, SLMB, PLUTO, which has an active area of 15.36×8.64 mm2, a pixel pitch size of 8 μm, and a total number pixels of 1920×1080) separately using 150mm lenses (L2, L3) with magnification of 15. The beam diameters at SLMA and SLMB are about 2mm. After transforming by SLMA and SLMB, the photons are collected into single mode fibers (SMF1, SMF2) using 4-f imaging systems with lenses groups (L4, L5 and L6, L7). The collected photons are detected using Si avalanche detector and free running InGaAs avalanche detector (IDQ 220), the output signals of the detectors are sent to coincidence device (Timeharp 260, pico, with coincidence window of 3.2ns).

The state of the generated non-degenerate two photon state correlated in OAM can be expressed as [33]

$$|\Phi\rangle = \sum_{m_s=-\infty}^{m_s=+\infty} c_{m_s} |m_s, \lambda_s\rangle |l_p - m_s, \lambda_i\rangle, \qquad (1)$$

Here $s$ and $i$ stand for the signal and idler respectively; $|m_s, \lambda_s\rangle$, $|l_p - m_s, \lambda_i\rangle$ denote the OAM eigenmodes of the signal and idler respectively; the OAMs carried by the signal ($m_s$) and idler ($l_p - m_s$) sum to the OAM carried by the pump beam $l_p$; $|c_{m_s}|^2$ is the probability to generate a photon pair with OAM of $m_s$, $l_p - m_s$; in the special case $l_p = 0$, for the symmetry of the SPDC process, we can infer that $c_{m_s} = c_{-m_s}$.

To show that the total OAM is conserved in the cascade processes of SFG and SPDC and the OAM information carried by the classical coherent light beams at 1550nm is transferred to the signal and idler photons at 795nma and 1550nm. We perform OAM correlation measurements for different signal and idler OAM. The experimental results for OAM of 0, -1 carried by the 1550nm coherent beam are showed in figure 2. Figure 2(a) is a three dimensional bar chart for signal and idler OAM ranges from -5 to 5 for pump beam's OAM of 0 . Due to misalignment, the coincidences of nearby OAMs are larger than non-adjacent OAMs. The maximum coincidence counts in 10 s are 31475 for signal and idler OAMs of 0, the smallest off-diagonal coincidence counts are about 100. Figure 2(b) is the corresponding coincidence counts for signal and idler OAM sum to 0. We can infer that the spiral bandwidth of this two color source is about 5. To increasing the spiral bandwidth, we can use a bigger beam waist, a shorter crystal length [34, 35] or change the phase matching condition by tuning the temperature the crystal [36]. Figure 2(b), (c) are the results for the pump OAM of -1, the coincidence counts when the signal and idler OAMs sum to -1 is much bigger than other uncorrelated cases. All these results are in agreements with equation (1), which imply that the total OAM is conserved in the cascade processes, and the OAM information is flowed from the classical party to the quantum party.

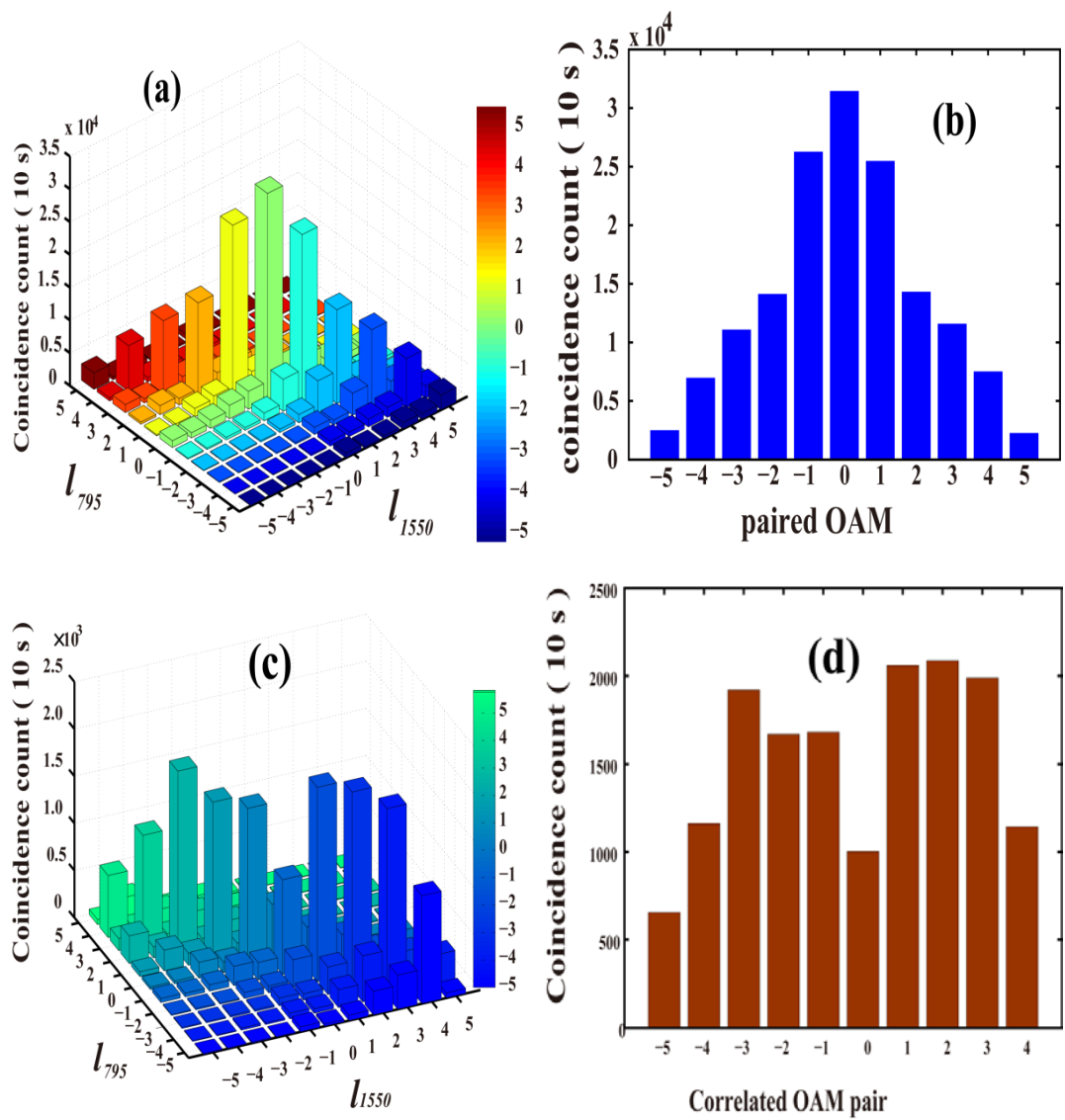

Figure 2. Experimental results for measuring the OAM correlation between signal an idler for different pump OAMs at 1550nm. (a), (c) show the coincidence counts in 10 s for signal and idler OAMs range from -5 to 5 for pump OAM of 0, -1; (b),(d)  shows the coincidence counts when the signal and idler OAMs sum to the pump OAM 0, -1 respectively.

Equation (1) also implies that the signal and idler OAMs are entangled in a high dimensional Hilbert space. To investigate the entanglement of the signal and idler photon, we simplify the measurements to 2-dimensional subspaces. For OAM entanglement in 2-dimensional subspaces, we can use Bloch-sphere to describe the state analogous to polarization states. An equally superposition of $|l\rangle$ and $|-l\rangle$ with arbitrary phase is represented by a point along the equator [37-39]. In an analogous fashion to the polarization states, we define the following superposition states

$$|\theta\rangle = \frac{1}{\sqrt{2}}\left(e^{il\theta}|l\rangle + e^{-il\theta}|-l\rangle\right), \tag{2}$$

These states are represented by points along the equator. States $|\theta\rangle$ are also called sector states, which have $2l$ sectors of alternating phases.

To quantify the entanglement, we must demonstrate that correlations between signal and idler are persist for superposition states. To verify this, we detect the photons in sector states defined in equation (2) oriented at different angles $\theta_A$ and $\theta_B$ respectively. Based on equation (1) and (2), the coincidence rate for detection one photon in $|\theta_A\rangle$ and the other in $|\theta_B\rangle$ is given by

$$C(\theta_A, \theta_B) = |\langle\theta_A|\langle\theta_B||\Phi\rangle|^2 \propto \cos^2[l(\theta_A - \theta_B)], \tag{3}$$

The high-visibility fringes of the joint probability are the signature of two-dimensional entanglement. Any modal impurities will degrade the quality of entanglement, which will reduce the visibilities of the fringes. By post-selection the state in equation (1) into two-dimensional subspaces, the states can be approximated as

$$|\Phi\rangle_l = \frac{1}{\sqrt{2}}(|l, 795\rangle|-l, 1550\rangle + |-l, 795\rangle|l, 1550\rangle), \tag{4}$$

The above states are normalized with the chosen subspaces and are entangled in OAM. By changing phase masks on SLMs, entangled states with any given $l$ can be prepared.

The experimental results in two-dimensional subspace of $l = \pm 1, \pm 2$ are showed in figure 3(a), (b). By fixing the orientation of the hologram in SLMB at angle $\theta_B = 0, \pi/4l$, we measure the coincidence count in 10 s by changing the angle from 0 to $\pi/l$. The angle period is $\pi/l$ depends on $l$. The visibilities of the sinusoidal fringes are 89.36%, 84.11% and 96.89%, 97.32% for $\theta_B = 0, \pi/4l$ ($l$=1, 2) respectively. All the visibilities are greater than 71%, which is sufficient to violate CHSH inequality and imply the present of two-dimensional entanglement. Therefore we further characterize the entanglement by measuring the CHSH inequality $S$ parameter [40, 41]. The definition of S in our experiment for angles $\theta_A$ and $\theta_B$ of the phase masks on the two SLMs are

$$S = E(\theta_A, \theta_B) - E(\theta_A, \theta'_B) + E(\theta'_A, \theta_B) + E(\theta'_A, \theta'_B), \tag{5}$$

The inequality is violated for values of $|S|$ which are greater than 2. $E(\theta_A, \theta_B)$ is calculated from the coincidence counts at particular sets of the hologram on the two SLMs,

$$E(\theta_A, \theta_B) = \frac{C(\theta_A, \theta_B) + C\left(\theta_A + \frac{\pi}{2l}, \theta_B + \frac{\pi}{2l}\right) - C\left(\theta_A + \frac{\pi}{2l}, \theta_B\right) - C\left(\theta_A, \theta_B + \frac{\pi}{2l}\right)}{C(\theta_A, \theta_B) + C\left(\theta_A + \frac{\pi}{2l}, \theta_B + \frac{\pi}{2l}\right) + C\left(\theta_A + \frac{\pi}{2l}, \theta_B\right) + C\left(\theta_A, \theta_B + \frac{\pi}{2l}\right)}, \tag{6}$$

For entanglement state $|\Phi\rangle_l$, the inequality is maximally violated for example when $\theta_A = 0$, $\theta_B = \pi/8l$, $\theta'_A = \pi/4l$ and $\theta'_B = 3\pi/8l$. The measured S parameters for $|\Phi\rangle_1$, $|\Phi\rangle_2$ are 2.28±0.016, 2.82± 0.0021 with 17 and 390 standard deviations respectively.

To completely describe an entangled state, reconstruction the density matrix is necessary. To reconstruct the density matrix of $|\Phi\rangle_l$, we follow the quantum tomography methods analogous to two-bit polarization entanglement states [42]. We define $|L\rangle = |l\rangle$ and $|R\rangle = |-l\rangle$. At least 16 projection measurements at the projection basis $|L\rangle, |R\rangle, 1/\sqrt{2}(|L\rangle+|R\rangle)$ and $1/\sqrt{2}(|L\rangle-i|R\rangle)$ are required. As a example, the results for state $|\Phi\rangle_2$ are showed in Figure 3(c), (d). The real and imaginary parts of the density matrix reconstruct using maximum-likelihood are showed in Figure 3(c), (d). The fidelity $F = \langle\Phi|_2 \rho_e |\Phi\rangle_2$ of the experimental reconstructed density matrix with respect to the theoretical density matrix is $0.84 \pm 0.0062$. The discrepancy in the fidelity is due to slightly different coupling efficiency between states $|L\rangle, |R\rangle$ and $1/\sqrt{2}(|L\rangle+|R\rangle)$, $1/\sqrt{2}(|L\rangle-i|R\rangle)$.

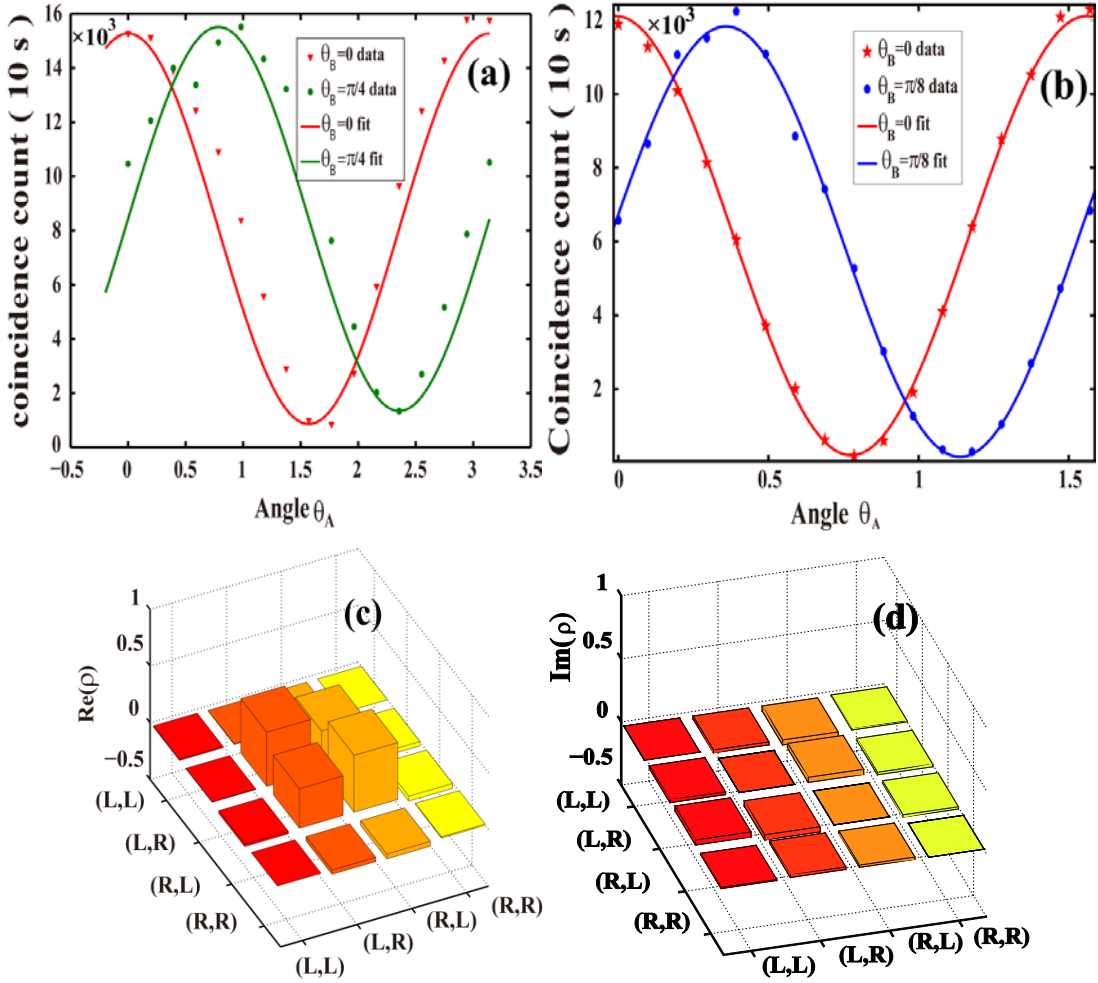

Figure 3. Experimental results for characterizing the entanglement in two-dimensional subspaces. (a), (b) show coincidence counts of signal and idler as a function of the phase mask angle $\theta_A$ when the phase mask angle $\theta_B = 0, \pi/4l$ ($l$=1, 2); (c), (d) are the real and imaginary parts of the reconstructed density matrix for $|\Phi\rangle_2$.

**Conclusion and discussion**

We have demonstrated a proof of principle experiments for building a link for classical telecom-communication network with a quantum network in OAM degree of freedoms of light based on cascade second order nonlinear frequency conversion. An OAM carried beam at telecom wavelength 1550nm is up-converted to OAM-carried beam at 525.5nm based on cavity enhanced SFG, then the up-converted OAM carried beam is used to pump a second crystal for SDPC to generate non-degenerate two-color OAM entangled signal and idler photon at 795nm and 1550nm. The dependence of the OAM correlation on the classical beam at 1550nm is observed, when the OAM carries by the classical beam is switched, the OAM correlation between signal and idler is also changed because of the total OAM is conserved in the cascade processes. Therefore the OAM information carries by the classical beam is transferred to the signal and idler. This two-color OAM entangled source is very promise in quantum communication, as one of the wavelength is corresponding to atom quantum memory of $Rb^{85}$ (the bandwidth can be narrowed using a cavity), the other wavelength is at telecom band which is suitable for long distance transmission. Our work is helpful in building a compatible interface between classical optical communication network and quantum communication network in OAM degree of freedoms, this hybrid network will be of both high capacity and superior security.

**Method**

**SFG cavity design.** The SFG cavity consists of 4 cavity mirrors (M1-M4), which is designed for single resonance at 795nm. The total length of the cavity is 547mm. The input coupling mirror M1 has a transmittance of 3% at 795nm. Mirror M2 is high reflective coated at 795nm (R >99.9%), a PZT is attached to it to scan and lock the cavity. The two concave mirrors M3 and M4 have curvature of 80mm, M3 is high transmittance coated at 1560nm (T>99%) and high reflectively coated at 795nm(R>99.9%), M4 is high transmittance coated at 525nm (T>98%) and high reflectively coated at 1560nm and 795nm (R>99.9%). The fundamental cavity mode has a beam waist of 33 μm at the middle of M4.

**Details of the SFG and SPDC crystals.** The type-I SFG and SPDC crystals are manufactured by Raicol crystals, they all have dimensions of 1mm×2mm×10mm. Their poling period is 9.375μm, both end faces of crystals are anti-reflected coated for 525nm, 795nm and 1560nm, the measured quasi-phasing matching temperature of PPKTP1 and PPKTP2 are 39.4℃ and 43.0℃ respectively.

**Holograms used in the experiments.** We use OAM index of 2 as an example to illustrate the holograms used for each measurement in the experiments. The holograms used are listed in Figure 4. Figure 4(c), (c) are the definitions of angles of $\theta_A$ and $\theta_B$ OAM index of 2, which are used in measurements in Figure 1. Figure 4(c), (d) shows the definition of on SLMs A and B for measurements the Figure 3(a), (b). Figure 4(e)-(h) are 4 orthogonal projection basis for the quantum tomography measurements in Figure 3(c), (d).

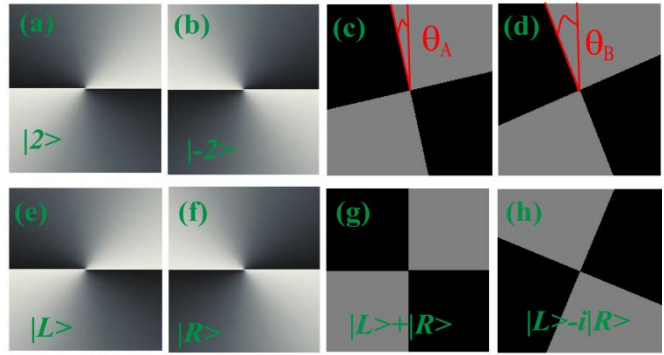

Figure 4. Holograms used in our experimental measurements.(a),(b) are the holograms for single OAM states used in measurements of Figure 1; (c), (d) are the definition of $\theta_A$ and $\theta_B$; (e)-(h) are holograms for quantum tomography measurements.